\def\papertitle{Audio-to-Audio via Diffusion Warm Initialization}
\def\paperauthorA{Cristóbal Andrade}
\def\paperauthorB{Sebastian J. Schlecht}

\documentclass[twoside,a4paper]{article}
\usepackage{etoolbox}

\usepackage[print]{dafx26v3}

\usepackage{amsmath,amssymb,amsfonts,amsthm}
\usepackage{siunitx}
\usepackage{euscript}
\usepackage[T1]{fontenc}
\usepackage[utf8]{inputenc}
\usepackage{ifpdf}
\usepackage[english]{babel}
\usepackage{caption}
\usepackage{subfig} 
\usepackage{color}
\usepackage{booktabs}
\usepackage{lipsum}
\usepackage{algorithm}
\usepackage{algpseudocode}

\input glyphtounicode
\ninept

\newcounter{numauth}
\newcounter{listcnt}
\newcommand\authcnt[1]{\ifdefined#1 \stepcounter{numauth} \fi}

\newcommand\addauth[1]{
\ifdefined#1 
\stepcounter{listcnt}
\ifnum \value{listcnt}<\value{numauth}
\appto\authorslist{, #1}
\else
\appto\authorslist{~and~#1}
\fi
\fi}
\authcnt{\paperauthorB}
\authcnt{\paperauthorC}
\authcnt{\paperauthorD}
\authcnt{\paperauthorE}
\authcnt{\paperauthorF}
\authcnt{\paperauthorG}
\authcnt{\paperauthorH}
\authcnt{\paperauthorI}
\authcnt{\paperauthorJ}
\def\authorslist{\paperauthorA}
\addauth{\paperauthorB}
\addauth{\paperauthorC}
\addauth{\paperauthorD}
\addauth{\paperauthorE}
\addauth{\paperauthorF}
\addauth{\paperauthorG}
\addauth{\paperauthorH}
\addauth{\paperauthorI}
\addauth{\paperauthorJ}

\usepackage{times}

\newif\ifpdf
\ifx\pdfoutput\relax
\else
   \ifcase\pdfoutput
      \pdffalse
   \else
      \pdftrue
   \fi
\fi

\ifpdf 
  \usepackage[pdftex,
    pdftitle={\papertitle},
    pdfauthor={\authorslist},
    pdfsubject={Proceedings of the 29th International Conference on Digital Audio Effects (DAFx26)},
    colorlinks=false, 
    bookmarksnumbered, 
    pdfstartview=XYZ 
  ]{hyperref}
  \usepackage[pdftex]{graphicx}
\else 
  \usepackage[dvips]{epsfig,graphicx}
  \usepackage[dvips,
    pdftitle={\papertitle},
    pdfauthor={\authorslist},
    pdfsubject={Proceedings of the 29th International Conference on Digital Audio Effects (DAFx26)},
    colorlinks=false, 
    bookmarksnumbered, 
    pdfstartview=XYZ 
  ]{hyperref}
\fi
\usepackage[hypcap=true]{caption}
\title{\papertitle}

\affiliation{
 \paperauthorA\, and \paperauthorB \,}
 {\href{https://www.lms.tf.fau.eu/}{Multimedia Communications and Signal Processing},\\ Friedrich-Alexander-Universität Erlangen-Nürnberg\\ Erlangen, Germany\\
 {\tt \href{mailto:cristobal.andrade@fau.de, sebastian.schlecht@fau.de}{\{cristobal.andrade, sebastian.schlecht\}@fau.de}}
 }

\begin{document}
\ifpdf 
  \DeclareGraphicsExtensions{.png,.jpg,.pdf}
\else  
  \DeclareGraphicsExtensions{.eps}
\fi


\maketitle

\begin{abstract}
In this paper, we propose diffusion warm initialization as a simple yet effective approach for a range of audio-to-audio transformation tasks. To illustrate the generality of the approach, we demonstrate its use in timbre transfer, MIDI-to-Real synthesis, and multiple audio enhancement tasks. We conduct a detailed empirical analysis on timbre transfer to investigate the role of the initialization time $t_\text{init}$. The effect of $t_\text{init}$ is evaluated using pitch-based Jaccard Distance and Fréchet Audio Distance to quantify faithfulness to the input signal and alignment with the target distribution.  Our results provide practical guidance for selecting $t_\text{init}$ and show that, once properly chosen, a single pretrained diffusion model combined with warm initialization can support multiple transformation objectives without task-specific training or conditioning. Despite its simplicity, this approach already achieves competitive results when compared with more complex pipelines designed specifically for these tasks. We further observe that warm initialization does not necessarily require explicit noise injection, as the guide signal itself can often serve as a valid initialization state for the backward diffusion process. Together, these findings show that warm initialization provides a simple and effective framework that serves as a fundamental building block for more complex audio transformation pipelines.

\end{abstract}

\section{Introduction}
\label{sec:intro}
Diffusion models are a class of generative models that have gained significant attention in recent years, with applications ranging from image synthesis \cite{rombach2022highresolutionimagesynthesislatent} and audio generation \cite{evans2024stableaudioopen} to protein structure design \cite{Denovodesignofprotein}. They operate by progressively corrupting data samples through a forward process that adds noise in small increments, while a neural network is trained to approximate the reverse process and remove this noise step by step \cite{ho2020denoisingdiffusionprobabilisticmodels}. In this way, the model learns to transform samples drawn from a Gaussian distribution, which is simple to sample from, into samples from a complex data distribution.

Warm initialization \textit{hijacks} the generative process by starting the reverse diffusion from a guide signal $\mathbf{x}^{(\mathrm g)}$, typically with noise added according to the chosen starting diffusion time, rather than from pure noise. As a result, the reverse diffusion process modifies an existing signal instead of synthesizing one entirely from scratch \cite{meng2022sdeditguidedimagesynthesis, kawar2023imagic, lee2026diffusiontimbretransfermutual}.
Such a mechanism is particularly suitable for audio-to-audio transformations, where certain characteristics of the input signal must be preserved while others are modified.

In this work, we show that warm initialization provides a simple yet effective method for a wide range of audio transformation tasks. To illustrate its versatility, we present examples including timbre transfer, MIDI-to-Real Synthesis and audio enhancement.  In this context, audio transformations via warm initialization can be interpreted as moving a sample $\mathbf{x}^{(\mathrm g)}$ from a low-probability region toward a higher-probability region under the data distribution $p_0$. Audio examples are provided on the companion website\footnote{\href{https://cristobalandrade.github.io/Audio-to-Audio-via-Diffusion-Warm-Initialization/}{cristobalandrade.github.io/Audio-to-Audio-via-Diffusion-Warm-Initialization}\label{fn:AudioExamples}}.

This work further explores the idea that unconditional generation with diffusion models can be repurposed for multiple tasks without task specific retraining \cite{moliner2023solvingaudioinverseproblems}.   In this context, we argue that warm initialization plays an important role, as it allows the reverse diffusion process to start in a neighborhood of the desired portion of the data distribution rather than from pure noise. By initializing the process with a meaningful guide signal $\mathbf{x}^{(\mathrm g)}$, the generation begins closer to signals that already contain relevant structure, enabling the model to modify selected characteristics while preserving others. We therefore propose warm initialization as a fundamental step in audio transformation pipelines.

Selecting the initialization time $t_\text{init}$ in warm initialization remains an open problem with limited theoretical grounding. Prior work such as AudioLDM \cite{liu2023audioldm} employs warm initialization for audio manipulation but does not provide guidance on how to select the initialization timestep.  In practice, this parameter is typically selected through manual tuning, balancing heuristic trade offs between realism and faithfulness \cite{meng2022sdeditguidedimagesynthesis}. Realism refers to how closely the output resembles a valid sample from the target data distribution, while faithfulness refers to how well the transformation preserves the required properties of the guide signal. In this work, we provide a systematic analysis of the realism–faithfulness tradeoff in audio warm initialization, introducing complementary audio-domain metrics to identify a suitable operating point for $t_\text{init}$, an aspect absent from prior work on audio warm initialization.

We also provide empirical evidence that explicit noise corruption of the guide signal $\mathbf{x}^{(\mathrm g)}$ is not strictly necessary for warm initialization. Our experiments show that using the guide signal directly as the starting point of the reverse diffusion can be sufficient, and that additional noise injection may introduce detrimental artifacts, such as generative hallucinations.

This paper is organized as follows. Section~\ref{background} introduces diffusion denoising models and the concept of warm initialization. Section~\ref{Method} describes the proposed framework and the metrics used to analyze and tune the initialization time. Section \ref{Experiments} presents experimental results and example applications. Section \ref{Discussion} discusses limitations and directions for future work. Finally, Section \ref{conclusion} provides concluding remarks.

\section{Diffusion Warm Initialization}
\label{background}
\subsection{Diffusion Models}
Diffusion models are a class of deep generative models that achie\-ve high sample quality with strong mode coverage \cite{xiao2022tacklinggenerativelearningtrilemma}. They learn to generate data by learning to transform a simple-to-sample distribution $p_T$, typically Gaussian noise $\mathbf{x}_T \sim \mathcal{N}(\mathbf 0, \mathbf I)$, to the data distribution $p_0$ \cite{ho2020denoisingdiffusionprobabilisticmodels}.

Training comprises a forward and a reverse process.  The forward process defines a discrete-time Markov chain defined over $t \in [0,T]$ that progressively corrupts a data sample $\mathbf{x}_0 \sim p_0$ according to a predefined noise schedule. Specifically, the transitions $q(\mathbf{x}_t \mid \mathbf{x}_{t-1})$ are Gaussian and chosen such that each latent variable admits the closed-form representation
$\mathbf{x}_t = \alpha_t\, \mathbf{x}_0 + \sigma_t \boldsymbol \varepsilon $, with  $\boldsymbol  \varepsilon \sim \mathcal{N}(\mathbf 0, \mathbf I)$,
where $\alpha_t$ and $\sigma_t$ are time-dependent coefficients controlling the signal and noise contributions, respectively. As $t$ increases, $\alpha_t$ decreases while $\sigma_t$ increases \cite{ho2020denoisingdiffusionprobabilisticmodels, song2022denoisingdiffusionimplicitmodels}. This construction induces a sequence of intermediate marginals $p_t$, with $\mathbf{x}_t \sim p_t$ that gradually transform the data distribution into Gaussian noise.

In the reverse process, a neural network $f_\theta$ is trained to approximate the reverse-time transition kernel of the forward Markov chain, namely the posterior distribution $q(\mathbf{x}_{t-1} \mid \mathbf{x}_t)$. In practice, this is achieved by learning a parameterization of the transition, commonly through prediction of the injected noise $\boldsymbol\varepsilon$ or the score function $\nabla_{\mathbf{x}_t} \log p_t(\mathbf{x}_t)$. These two parameterizations are mathematically linked via Tweedie’s formula \cite{efron2011tweedies}. By approximating the posterior transitions, the learned model progressively transforms samples from noise toward the data manifold.

After training, generation is performed by sampling $\mathbf{x}_T \sim \mathcal{N}(\mathbf 0, \mathbf{I})$ and iteratively applying the learned reverse transitions, transporting the sample through the intermediate distributions $p_t$ toward $p_0$. The final output of this process is denoted by $\hat{\mathbf{x}}_0$, corresponding to a generated sample from the model distribution approximating $p_0$. Under this formulation, if samples $\mathbf x_t \sim p_t$ were available, the reverse diffusion process could be initialized directly from these samples, thereby generating outputs that lie on  $p_0$.


\subsection{Warm Initialization}


Warm initialization is a technique in diffusion models in which the backward process is initialized from an intermediate diffusion timestep $t_\text{init} \in [0,T]$. In this work, we parameterize this choice using $\tau_{\text{init}} \in [0,1]$, which denotes the fraction of the reverse diffusion trajectory that is skipped. The reverse process is initialized with a guide signal $\mathbf{x}^{(\mathrm g)}$ from a distribution $p_\mathrm g \approx p_{t_\text{init}}$ that lies closer to the data distribution $p_0$ than a pure noise distribution $p_T$ \cite{scholzWarmStartsAccelerate2025}. It is commonly employed to accelerate sampling by truncating the reverse trajectory, or to incorporate structured guidance in enhancement tasks \cite{moliner2024buddysinglechannelblindunsupervised, moliner2024blindaudiobandwidthextension}, where the generation process should only modify degraded components while preserving the remaining characteristics of the input signal.

Warm initialization  can be expressed as:
\begin{equation}
    \mathbf{x}^{(\mathrm g)}_{t_\text{init}} \sim \mathcal{N}(\alpha_{t_\text{init}}\mathbf{x}^{(\mathrm g)}, \sigma^2_{t_\text{init}}\mathbf{I})
    \label{eq:warm_init}
\end{equation}
Two standard formulations are commonly found in the literature: Variance Exploding (VE) and Variance Preserving (VP). In the VE formulation, $\alpha_t = 1$ for all $t$, and $\sigma_T$ is selected sufficiently large such that the terminal marginal $p_T$ approaches $\mathcal{N}(0, \sigma_T^2 I)$. In contrast, in VP case the coefficient satisfy $\alpha_t^2 + \sigma_t^2 = 1$ for all $t$, with $\sigma_t$ decreasing to 0 as $t$ approaches $0$, resulting in $p_T = \mathcal{N}(\mathbf 0, \mathbf I)$.  In this work, we use VE as it has demonstrated good empirical performance in prior studies \cite{meng2022sdeditguidedimagesynthesis}.

The central assumption underlying warm initialization is that the injected Gaussian noise $\mathcal{N}(\mathbf 0, \sigma_{t_\text{init}}^2 \mathbf I)$ should be chosen such that the noised guide $\mathbf{x}^{(\mathrm g)}_{t_\text{init}}$ lies in a high-density region of an intermediate marginal of the reverse diffusion process \cite{meng2022sdeditguidedimagesynthesis}. In our experiments, we show that explicit addition of Gaussian noise to the guide signal is not always necessary, as $\mathbf{x}^{(\mathrm g)}$ can already be treated as a sample from an intermediate marginal $p_t$, allowing initialization without additional noise injection.

To the best of our knowledge, there is a lack of theoretical evidence for systematically selecting the initialization time $t_\text{init}$. Existing studies provide partial insights into the role of initialization but do not offer practical guidance. For example, \cite{chung2022comecloserdiffusefasteracceleratingconditionaldiffusion} shows that the number of required reverse diffusion steps decreases as the initialization approaches the data distribution $p_0$, motivating starting closer to the data manifold. Similarly, image editing methods such as SDEdit \cite{meng2022sdeditguidedimagesynthesis} observe that the discrepancy between the guide signal and the generated output decreases as the initialization point moves closer to the end of the reverse diffusion process.

In practice, $t_\text{init}$ is typically selected heuristically through manual tuning. A commonly discussed heuristic is the trade off between realism and faithfulness. Earlier initialization encourages stronger projection onto the learned data manifold, often producing more realistic outputs but increasing deviation from the guide signal. Conversely, later initialization preserves structural properties of $\mathbf{x}^{(\mathrm g)}$ while limiting the extent of model driven modification \cite{meng2022sdeditguidedimagesynthesis, moliner2024buddysinglechannelblindunsupervised}. In the following sections, we examine this behavior empirically and provide guidance for identifying this operating regime in the audio domain.


\section{Audio-to-Audio via warm initialization}
\label{Method}
In this section, we present the general framework of warm initialization for audio-to-audio tasks. We also introduce two metrics designed to assess the trade-off between realism and faithfulness and to help identify a suitable initialization time $t_\text{init}$.


\subsection{Framework}
\begin{algorithm}[!t]
\caption{Diffusion Warm Initialization}
\begin{algorithmic}[1]
\Require Input signal guide $\mathbf{x}^{(\mathrm g)}$, initialization fraction $\tau_\text{init}\in [0,1]$, desired noise portion $\lambda  \in [0,1]$, diffusion noise schedule $ \sigma_t$, diffusion denoiser $f_\theta$, number of inference steps $T$, guidance scale $\omega$
\State $t_\text{init} \gets \lceil (1 - \tau_\text{init})\cdot T \rceil$
\State Sample $\boldsymbol \varepsilon \sim \mathcal{N}(\mathbf 0, \mathbf I)$
\State $\mathbf{x}_{t_\text{init}} \gets  \mathbf{x}^{(\mathrm g)} + \lambda \,\sigma_{t_\text{init}}\, \boldsymbol \varepsilon$
\For{$t \gets t_\text{init}$ \textbf{down to} $1$}
    \State $\mathbf{x}_{t-1} \gets \text{Diffusion Pipeline Step}(\mathbf{x}_t, t; f_\theta, \omega)$
\EndFor
\State \Return $\mathbf{x}_0$
\end{algorithmic}
\label{alg:warm_init}
\end{algorithm}

Algorithm \ref{alg:warm_init} provides a unified formulation of diffusion warm initialization as a modification of the standard diffusion sampling procedure for audio transformation tasks, where the reverse trajectory is initialized from a guide signal $\mathbf{x}^{(\mathrm g)} $.

\begin{figure*}[!h]
    \centering
    \includegraphics[width=\linewidth]{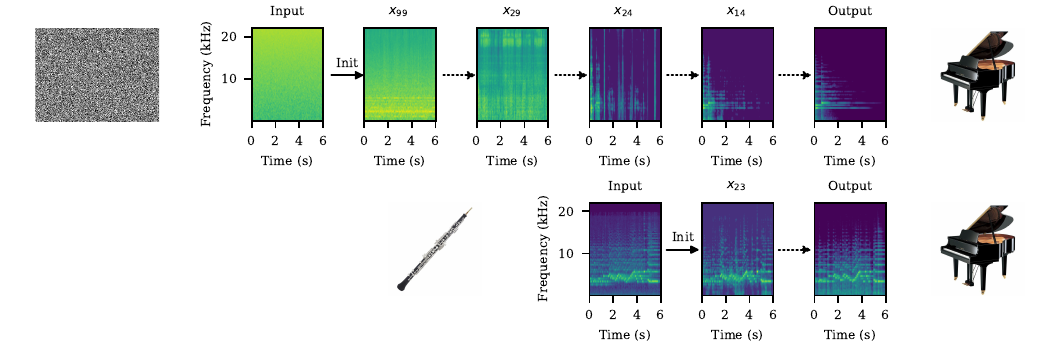}
    \caption{Diffusion generation compared with warm initialization. Dashed arrows indicate reverse diffusion steps. Top: Standard generation from Gaussian noise gradually produces piano audio. Bottom: Warm initialization from $\mathbf{x}^{(\mathrm g)}$ (Oboe) preserves the melody while the diffusion process shifts the timbre toward the piano distribution.}
    \label{fig:warm_init_vs_generation}
\end{figure*}

Figure \ref{fig:warm_init_vs_generation} illustrates the difference between a standard unconditional generative pipeline and a warm initialization pipeline in the context of timbre transfer. In this setting, the target data distribution $p_0$ corresponds to piano recordings. In the unconditional case, generation starts from Gaussian noise and progressively converges toward samples from $p_0$. In contrast, warm initialization starts from a guide signal $\mathbf{x}^{(\mathrm g)}$, here an oboe recording, and applies reverse diffusion to transform its timbre while preserving its musical structure.

For the selected $t_\text{init}$ in Figure \ref{fig:warm_init_vs_generation} the melody and temporal structure of the oboe signal are retained, while its spectral characteristics evolve toward those of a piano. This can be observed in the spectrogram, where the harmonic content shifts from an oboe-like distribution to a piano-like distribution. Importantly, the model is not conditioned on this specific oboe sample and may not have seen the underlying melody during training, yet it is able to map the signal toward the piano distribution $p_0$ through reverse diffusion.

This example highlights how warm initialization enables controlled transformations: instead of generating from noise, the model starts from a structured input and selectively modifies its characteristics. A more detailed analysis of timbre transfer using warm initialization is provided in Section \ref{Experiments}.

\begin{figure}[ht]
    \centering
    \includegraphics[width=\linewidth]{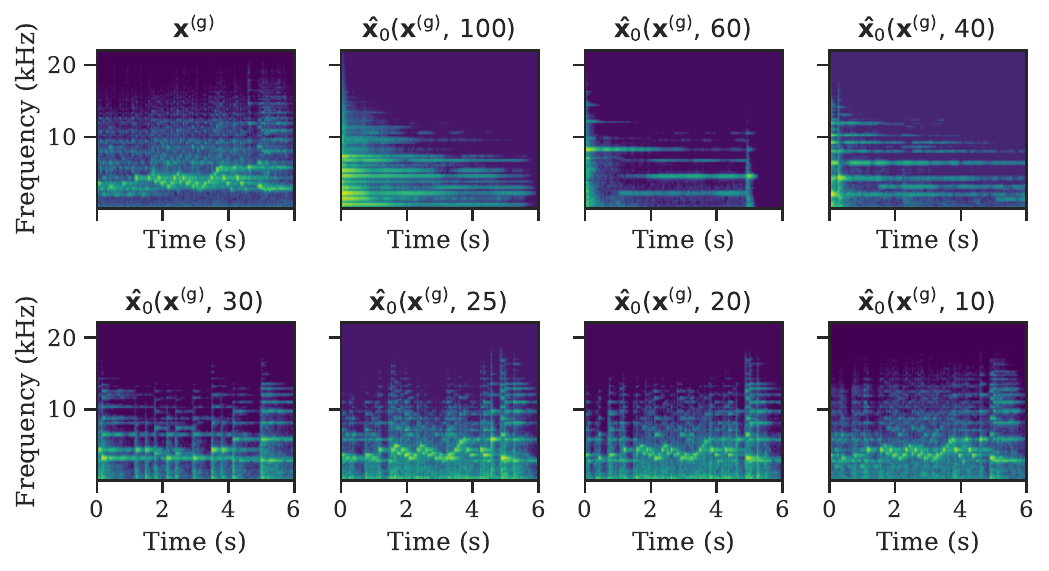}
    \caption{Spectrograms of oboe-to-piano timbre transfer for different initialization times $t_{\text{init}}$. A diffusion schedule with $T = 100$ steps is used, with reverse diffusion starting at $t_{\text{init}}$. Smaller $t_{\text{init}}$ leads to stronger deviation from the input, while larger values preserve more of the original melodic structure.}
    \label{fig:init_time_spectrograms}
\end{figure}

The choice of the initialization time $t_\text{init}$ determines how strong\-   ly the diffusion process modifies the guide signal. 
Figure \ref{fig:init_time_spectrograms} shows the spectrograms of the outputs $\hat{\mathbf{x}}_0(\mathbf{x}^{(g)}, t_{\text{init}})$ obtained via reverse diffusion for different values of $t_{\text{init}}$, using $T = 100$ inference steps. Earlier initialization results in the model modifying the signal more aggressively, producing sounds that better match the target distribution but preserve less of the guide’s structure. Later initialization preserves more characteristics of the input spectrogram while limiting the extent of the modification\footref{fn:AudioExamples}.

The parameter $\lambda$ in Algorithm \ref{alg:warm_init} controls the portion of the noise schedule used for initialization. In the following sections, we show empirically that adding Gaussian noise is not always necessary, and that $\lambda = 0$ often yields better results. This suggests that the guide signal $\mathbf{x}^{(\mathrm g)}$ may already lie close to an intermediate marginal $p_t$, allowing the reverse diffusion process to be initialized without additional noise. Increasing $\lambda$ can be detrimental, as additional noise injection tends to introduce hallucinated artifacts in the generated signal.

For this framework, we use the Stable Audio Open diffusion model \cite{evans2024stableaudioopen}. We use text conditioning to steer generation toward the desired target distribution. Conditioning is implemented via classifier-free guidance (CFG), where the denoiser combines unconditional and text-conditional predictions, and a guidance scale $\omega$ controls the strength of this signal \cite{hoClassifierFreeDiffusionGuidance2022}. For instance, in timbre transfer experiments targeting piano, we employ the prompt \textit{``Grand Piano, Chords and Melodies’’}. The reverse diffusion is initialized from the guide signal $\mathbf{x}^{(\mathrm g)}$, which defines the starting point of the trajectory, while the text prompt guides the subsequent evolution toward the target distribution. The guidance scale $\omega$ is included in Algorithm \ref{alg:warm_init}.

When applied in the context of warm initialization, we observe that the model requires a substantially higher guidance scale $\omega$ than the training setting, where $\omega =7$ was used.  In practice, this corresponds to increasing the CFG strength, thereby amplifying the conditional signal and improving fidelity to the target distribution \cite{hoClassifierFreeDiffusionGuidance2022}. This limitation would not arise in a model trained exclusively on the target domain, such as a piano-only generator, where alignment with the target distribution is an intrinsic property of the model.

 \subsection{Metrics}
 \label{Metrics}
To assess the trade-off between alignment with the target distribution and preservation of the characteristic structure of the input signal, we employ two complementary metrics: Fréchet Audio Distance (FAD), which measures perceptual similarity in a learned embedding space, and Jaccard distance over pitch sets, for quantifying melodic consistency.

\subsubsection{Fréchet Audio Distance (FAD)}
 The Fréchet Audio Distance (FAD) measures how close a reference and a test set of audio samples are  in an embedding space and it is often used as a proxy for generation quality \cite{kilgour2019frechetaudiodistancemetric}.
For its computation both the reference and test audio embeddings are modeled with multivariate normal distributions. Let $\mu_r$ and $\mu_t$ denote the means, and $\Sigma_r$ and $\Sigma_t$ the covariance matrices, of the reference and test distributions, respectively. The FAD between two gaussian is then given by:    \begin{equation}
        \mathrm{FAD} = \lVert \mu_r - \mu_t \rVert^2 
        + \operatorname{tr}\left( 
        \Sigma_r + \Sigma_t - 2\sqrt{\Sigma_r \Sigma_t} 
        \right)
    \end{equation}
In this work we use the toolbox provided by \cite{guiAdaptingFrechetAudio2024} using the \textit{LAION-CLAP} embedding as it is used in Stable Audio \cite{evans2024stableaudioopen}.

\subsubsection{Jaccard Distance (JD)}

To assess the faithfulness of the generated output with respect to the guide signal, we compute the Jaccard Distance (JD)
    \begin{equation}
        \mathrm{JD}(A, B) = 1 - \frac{|A \cap B|}{|A \cup B|}
    \end{equation}
where A and B are sets of pitch values extracted from the guide signal $\mathbf{x}^{(g)}$
and the generated output $\hat{x}_0$
respectively. The pitch sets are obtained using the mono pitch
between the reference and test melodies using the mono pitch algorithm based on MELODIA \cite{6155601} as implemented in the Essentia library \cite{10.1145/2502081.2502229}. Lower JD values indicate higher melodic similarity between the generated output and the guide signal $\mathbf{x}^{(\mathrm g)}$.

\section{Applications}
\label{Experiments}
In this section, we demonstrate the versatility of warm initialization across several audio-to-audio tasks, including timbre transfer, MIDI-to-real synthesis, and audio enhancement. As the main experimental study, we analyze timbre transfer in detail to investigate the role of the initialization time $t_\text{init}$ and its effect on the trade-off between faithfulness to the input signal and alignment with the target distribution. The remaining applications illustrate how the same framework can be applied to different tasks without task-specific modifications.

\subsection{Timbre Transfer}
Timbre denotes the perceptual properties that distinguish two instruments playing the same note at the same intensity, beyond pitch and amplitude. Timbre transfer refers to altering the instrument identity of an audio signal while preserving melody, rhythm, and other musical structure.

Several deep learning approaches have been proposed for this task. These include VAE-based methods \cite{esling2018generativetimbrespacesregularizing, caillon2021ravevariationalautoencoderfast, colonel2020conditioningautoencoderlatentspaces}, DDSP-based approaches \cite{carney2021tonetransfer}, and diffusion-based models. Diffusion approaches either condition generation on a target timbre \cite{ baoueb2024wavetransferflexibleendtoendmultiinstrument, comanducci2023timbretransferusingimagetoimage, garcía2025sketch2soundcontrollableaudiogeneration, li2024musicstyletransfertimevarying, huang2024musicstyletransferdiffusion} or bridge separately trained source and target models through forward–reverse diffusion schemes \cite{mancusi2025latentdiffusionbridgesunsupervised}. In \cite{lee2026diffusiontimbretransfermutual}, warm initialization is presented as a \textit{naive} baseline, where $t_\text{init}$ is selected by training a classifier and choosing the time step at which the input is classified as noise with 50\% probability. Similarly, AudioLDM \cite{liu2023audioldm} proposes
warm initialization in the latent domain for style transfer, but does not provide guidance on how to select $t_\text{init}$.

\begin{figure}[ht]
    \centering
    \includegraphics[width=\linewidth]{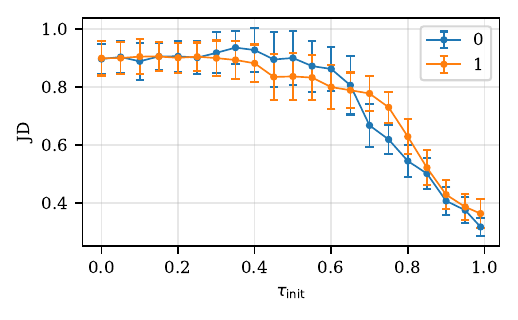}
    \caption{JD as a function of $\tau_{\text{init}}$ for oboe-to-piano timbre transfer for $\lambda =0$ and $\lambda =1$. $\tau_{\text{init}}=0$ denotes that the model performs all reverse steps, while $ \tau_{\text{init}}=1$ denotes that no reverse steps are performed.  Lower values indicate greater melodic similarity to the guide signal $\mathbf{x}^{(\mathrm g)}$.}
    \label{fig:jaccard_vs_t_Oboe2Piano}
\end{figure}

As a case study, we perform timbre transfer from oboe to piano using Algorithm~\ref{alg:warm_init} with $\lambda \in \{0,1\}$, where $\mathbf{x}^{(\mathrm g)}$ is an oboe sample obtained from MUSOPEN \cite{musopen_platform}. The diffusion model is run for $T=100$ inference steps with a guidance scale of $\omega =30$. To assess the effect of the initialization time $t_\text{init}$, we compute the JD between $\mathbf{x}^{(\mathrm g)}$ and 50 samples generated via warm initialization for each configuration, and report the mean and standard deviation. In addition, we compute FAD between a reference set of 100 unconditionally generated piano,  produced using the default guidance scale $\omega =7$, and 100 samples generated via warm initialization. The corresponding results are shown in  Figures~\ref{fig:jaccard_vs_t_Oboe2Piano} and \ref{fig:fad_vs_t_Oboe2Piano}, respectively.  The same analysis using FAD and JD is provided in the Appendix Section \ref{Appendix} for a string to clarinet timbre transfer experiment.

\begin{figure}[ht]
    \centering
    \includegraphics[width=\linewidth]{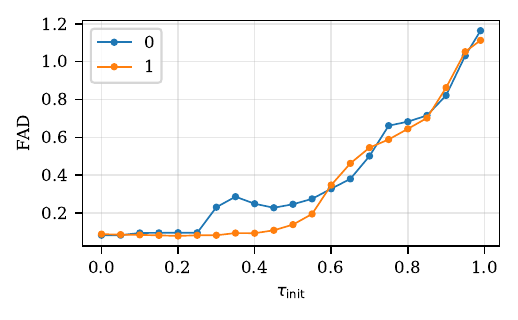}
    \caption{FAD as a function of $\tau_{\text{init}}$ for oboe-to-piano timbre transfer  for $\lambda =0$ and $\lambda =1$. $\tau_{\text{init}}=0$ denotes that the model performs all reverse steps, while $ \tau_{\text{init}}=1$ denotes that no reverse steps are performed. Lower values indicate closer alignment with the piano reference distribution.}
    \label{fig:fad_vs_t_Oboe2Piano}
\end{figure}

Figure \ref{fig:jaccard_vs_t_Oboe2Piano} shows that, for both values of $\lambda$, the melodic distance to the guide signal $\mathbf{x}^{(\mathrm g)}$ is initially high and decreases substantially once $\tau_\text{init}\gtrsim 0.65$. In contrast, Figure~\ref{fig:fad_vs_t_Oboe2Piano} indicates that later initialization leads to outputs that deviate further from the target piano distribution, as reflected by increasing FAD. We observe that the FAD is greater than zero at $\tau_\text{init}= 0$. This residual is a consequence of the high guidance scale used for the transfer, which differs from the scale applied during the unconditional generation of the reference samples.

For $\lambda = 1$, the JD remains consistently higher for  $t_\text{init}> 0.65$, while FAD values are largely comparable between $\lambda=0$ and $\lambda=1$. Informal listening suggests that the increased melodic deviation corresponds to stronger generative hallucinations, where the model introduces musical events not present in the guide signal. This indicates that stronger noise injection amplifies hallucinated content without substantially improving alignment with the piano distribution. Listening suggests the presence of a perceptual sweet spot around $\tau_\text{init}\approx 0.8$, indicating that a Jaccard distance below 0.6 and FAD below 0.7 may serve as a practical rule of thumb for selecting $\tau_\text{init}$.  It is worth noting that the location of the sweet spot may vary depending on the diffusion model used for generation.

Audio examples illustrating the effect of varying $\tau_\text{init}$ and the influence of $\lambda$ within the identified sweet spot are provided on the companion website. In addition to the oboe-to-piano experiments, we include further timbre transfer cases for comparison with prior work: string-to-clarinet in relation to \cite{baoueb2024wavetransferflexibleendtoendmultiinstrument, comanducci2023timbretransferusingimagetoimage}, synth-to-violin compared to \cite{lee2026diffusiontimbretransfermutual}, and violin-to-flute in comparison with \cite{mancusi2025latentdiffusionbridgesunsupervised}. Thereby we set $\lambda = 0$. These examples indicate that our method performs competitively despite its simpler formulation.

We also include further exploratory examples in which different input signals are transformed to piano. These include humming, nature recordings, and computer mouse clicks, which produce toy piano–like sounds, isolated piano notes, and plucked piano–string–like transients.

\subsection{MIDI-to-Real}
MIDI-to-Real synthesis refers to transforming audio rendered from symbolic MIDI into realistic recordings that capture timbral detail and characteristics of human performance. Several approaches found in literature build on Differentiable Digital Signal Processing (DDSP) \cite{engel2020ddspdifferentiabledigitalsignal}. For example, MIDI-DDSP employs a three-stage pipeline that maps MIDI notes to expressive performance features and subsequently synthesizes audio using a DDSP-based renderer \cite{wu2022mididdspdetailedcontrolmusical}. Castellon et al.~\cite{castellonRealisticMIDIInstrument} similarly train a neural network to predict synthesis parameters from MIDI, which are then used to drive a DDSP synthesizer. 

More recently, diffusion-based approaches have also been explored. Take et al.~\cite{take2025annotationfreemiditoaudiosynthesisconcatenative} propose a diffusion-based refinement me\-thod that operates on rendered audio in a manner related to our approach, although practical guidance for selecting the initialization time $t_\text{init}$ is not discussed.

In our experiments, we treat MIDI-to-Real synthesis as an audio-to-audio refinement task and apply warm initialization with $\lambda = 0$. Audio examples comparing our method with prior work are provided on the companion website.  From these examples, we observe that the strength of the transformation depends on the quality of the initial MIDI rendering. Despite its simplicity, the proposed framework is able to produce realistic outputs and achie\-ves performance comparable to more complex pipelines designed specifically for this task.
\subsection{Audio Enhancement}
Audio enhancement refers to the task of improving the quality of recorded audio signals, including speech, music, and impulse responses. Degradations may arise from environmental factors such as background noise or interference, as well as from limitations or suboptimal configurations of the recording hardware and acquisition system. Common enhancement tasks include dereverberation \cite{moliner2024buddysinglechannelblindunsupervised}, declipping \cite{moliner2023solvingaudioinverseproblems, lee2026solvingroomimpulseresponse} and denoising\cite{godsill1998digital}.

Historically, audio enhancement has been addressed using classical signal processing techniques \cite{godsill1998digital, NIPS2000_65699726}, and more recently through deep learning approaches \cite{yu2020speech, unknown}. In particular, diffusion-based methods have shown promising results for restoration and enhancement tasks \cite{moliner2024buddysinglechannelblindunsupervised,  kandpal2022musicenhancementimagetranslation, manor2024zeroshotunsupervisedtextbasedaudio}.

\begin{figure}[t]
    \centering
    \includegraphics[width=\linewidth]{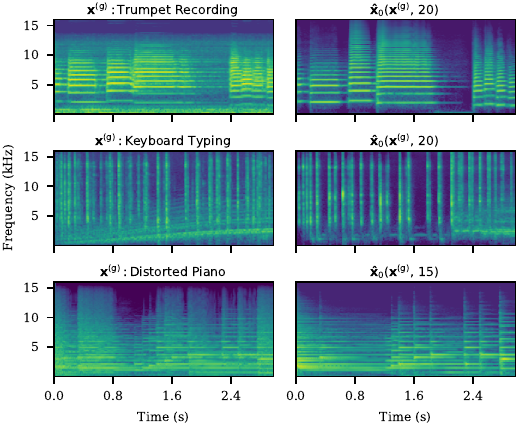}
    \caption{Spectrograms of enhanced audio signals. Left: degraded guide signals $\mathbf x^{(\mathrm g)}$. Right: corresponding outputs obtained via warm initialization.}
    \label{fig:enhancemnet_spectrograms}
\end{figure}

In our experiments, we apply warm initialization with $\lambda = 0$ to several enhancement scenarios, including declipping, denoising, and source interference suppression. Figure~\ref{fig:enhancemnet_spectrograms} shows a few examples, including a noisy trumpet recording, keyboard typing with a siren in the background, and a distorted piano signal. Audio examples comparing our approach with prior work are provided on the companion website. These examples show that the proposed framework can be applied to a range of enhancement tasks while remaining conceptually simple.

\section{Discussion}
\label{Discussion}
While warm initialization is a versatile method, it presents certain limitations. One limitation we encountered during experimentation is that the method shows reduced performance when the guide signal contains human voice, as in humming-to-piano or sketch-to-sound scenarios, where the resulting outputs exhibit lower quality\footref{fn:AudioExamples}. This suggests that the model fails to consistently move such inputs toward high-density regions of the target distribution during reverse diffusion. Further investigation is required to determine whether this behavior is inherent to the model or a consequence of the initialization strategy. 

Another limitation arises in source separation tasks. The method often fails to remove unwanted components; instead of suppressing them through reverse diffusion, these components tend to persist or are altered into more artificial or distorted versions. This behavior is particularly evident for highly percussive elements, such as drums, which are not effectively suppressed by the framework. This provides a clear case where additional mechanisms on top of warm initialization may be required to achieve selective suppression.

A further practical consideration concerns the sensitivity of the method to its hyperparameters. The guidance scale $\omega$, the noise injection parameter $\lambda$, and the initialization time $\tau_\text{init}$ all influence the output. In our experiments, the sweet spot around $\tau_\text{init} \approx 0.8$ generalizes across both timbre transfer experiments. However, these values are specific to Stable Audio Open. Different diffusion models have different probability paths and noise schedules, causing the operating regime for $\tau_\text{init}$ to shift accordingly.  The metric-based framework proposed in Section \ref{Metrics} provides a practical tool for re-tuning these parameters in such cases.

More broadly, the approach relies on empirical tuning of the initialization time $t_\text{init}$ and lacks a theoretical framework relating initialization to the structure of the data and marginal manifolds. These limitations highlight the need for further theoretical investigation of warm initialization.

\section{Conclusion}

\label{conclusion}
In this work, we investigated diffusion warm initialization as a simple framework for audio-to-audio tasks. We showed that warm initialization enables a single pretrained diffusion model to support a range of transformations, including timbre transfer, MIDI-to-real synthesis, and audio enhancement, without task-specific retraining or conditioning. 

A detailed empirical study on timbre transfer examined the role of the initialization time $t_\text{init}$. Using Fréchet Audio Distance and pitch-based Jaccard distance, we analyzed the trade-off between alignment with the target distribution and preservation of the guide signal. The results reveal the presence of a practical operating regime for $t_\text{init}$, providing empirical guidance for selecting this parameter in audio transformation tasks.

We further observed that explicit corruption of the guide signal with Gaussian noise is not always required. In many cases, the guide signal itself can already lie close to an intermediate diffusion state, allowing the reverse process to start directly from the input signal. Additional noise injection may instead introduce hallucinated content and reduce fidelity to the guide.

Our findings suggest that warm initialization constitutes a simple but effective mechanism for adapting pretrained diffusion models to audio-to-audio tasks. Future work may investigate the theoretical properties of this mechanism and explore how the latent structure of diffusion models influences the transformations induced by warm initialization.


\bibliographystyle{IEEEtranDAFx}
\bibliography{DAFx26_tmpl} 

\section{Appendix: String-to-Clarinet Timbre Transfer}
\label{Appendix}
\begin{figure}[ht]
    \centering
    \includegraphics[width=\linewidth]{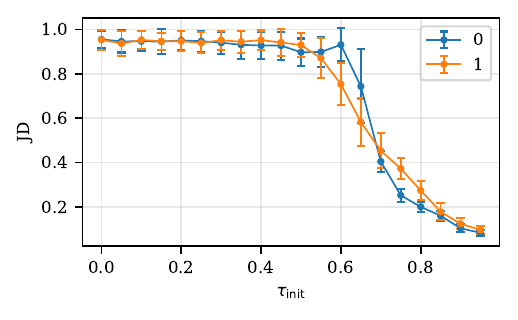}
    \caption{JD as a function of $\tau_{\text{init}}$ for string-to-clarinet timbre transfer  for $\lambda =0$ and $\lambda =1$. $\tau_{\text{init}}=0$ denotes that the model performs all reverse steps, while $ \tau_{\text{init}}=1$ denotes that no reverse steps are performed.}
    \label{fig:jaccard_vs_t_mono_melody_String2Clarinet}
\end{figure}

\begin{figure}[ht]
    \centering
    \includegraphics[width=\linewidth]{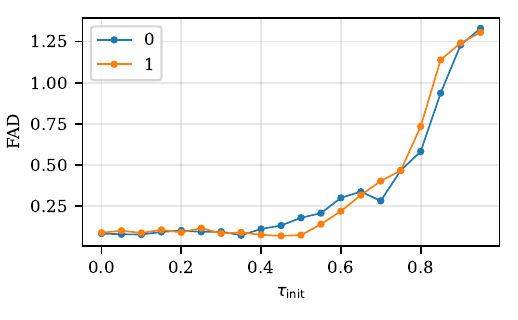}
    \caption{FAD as a function of $\tau_{\text{init}}$ for string-to-clarinet timbre transfer  for $\lambda =0$ and $\lambda =1$. $\tau_{\text{init}}=0$ denotes that the model performs all reverse steps, while $ \tau_{\text{init}}=1$ denotes that no reverse steps are performed.}
    \label{fig:fad_vs_t_String2Clarinet}
\end{figure}

Similar to the oboe-to piano experiment, we perform timbre transfer from string-to-clarinet using Algorithm \ref{alg:warm_init}. The diffusion model is run for T = 100 inference steps with a guidance scale of $\omega = 30$. To evaluate the effect of the initialization time $t_\text{init}$, we compute the Jaccard Distance (JD) between the guide signal and 50 samples generated via warm initialization for each value of $t_\text{init}$, and report the mean and standard deviation. In addition, we compute FAD between a reference set of 100 unconditionally generated clarinet samples, produced using the default guidance scale $\omega=7$, and 100 samples generated via warm initialization. The corresponding results are shown in Figures~\ref{fig:jaccard_vs_t_mono_melody_String2Clarinet} and~\ref{fig:fad_vs_t_String2Clarinet}.

Informal listening suggests a perceptual sweet spot around $\tau_\text{init}= 0.8$. This is consistent with the guideline observed in the oboe-to-piano experiment, where a sweet spot tends to appear once the Jaccard distance is below 0.6 and FAD is below 0.7. In this case, FAD acts as the limiting factor. Audio examples are available on the companion website.

\end{document}